\documentclass[referee]{aa}         
\usepackage{graphicx}
\usepackage[]{natbib}
\usepackage{txfonts}
\usepackage{color}
\newcommand{\Msun}{$\rm M_{\odot}$}
\newcommand{\Zsun}{$\rm Z_{\odot}$}
\newcommand{\Lsun}{\mbox{$L_{\rm sun}$}}

\newcommand{\Teff}{\mbox{$T_{\rm eff}$}}

\newcommand{\logg}{\mbox{$\log$~\textsl{g}}~}

\newcommand{\Mdot}{$\dot M$}

\newcommand{\kms}{km s$^{-1}$}

\begin{document}
   \title{OB stars at the lowest Local Group metallicity}

\subtitle{GTC-OSIRIS observations of Sextans A
 \thanks{
 Based on observations made with the Gran Telescopio Canarias (GTC), 
 installed in the Spanish Observatorio del Roque de los Muchachos 
 of the Instituto de Astrof\'{\i}sica de Canarias, on the island of La Palma.
 Programme ID GTC59-12A.}
       }
   \author{I. Camacho\inst{1,2}
          \and
          M. Garcia\inst{3}
          \and
          A. Herrero\inst{1,2}
          \and
          S. Sim\'on-D\'{i}az\inst{1,2}
          }

   \institute{Instituto de Astrof\'{i}sica de Canarias, C/ V\'{i}a L\'{a}ctea s/n E-38205 La Laguna, Tenerife, Spain.\\
              \email{icamacho@iac.es}
         \and
             Departamento de Astrof\'{i}sica, Universidad de La Laguna, Avda. Astrof\'{i}sico
                Francisco S\'anchez, s/n, E-38200 La Laguna, Tenerife, Spain. \\
          \and
          Centro de Astrobiolog\'\i a (CSIC/INTA), Instituto Nacional de T\'ecnica Aeroespacial, Ctra de Torrej\'on de Ajalvir, km 4, 28850 Torrej\'on de Ardoz, Madrid, Spain.          
             }

   \date{Received xxx, 201x; accepted xxxx, 201x}

 
  \abstract
 {Massive stars have an important role in the chemical and dynamical evolution of the Universe. The first metal-poor stars may have started the reionization of the Universe. In order to understand these early epochs it is necessary to know the behavior and the physical properties of massive stars in very metal-poor environments. We focus on the massive stellar content of the metal-poor irregular galaxy Sextans A.}
   {Our aim is to find and classify OB stars in Sextans A, to later determine accurate stellar parameters of these blue massive stars in this low metallicity region $(Z \sim 0.1 \rm Z_{\odot})$.}
 {Using UBV photometry, the reddening-free index Q and GALEX imaging, we built a list of blue massive star candidates in Sextans A. We obtained low resolution (R $\sim$ 1000) GTC-OSIRIS spectra for a fraction of them and carried out spectral classification. For the confirmed O-stars we derive preliminary stellar parameters.}
{The target selection criteria and observations were successful and have produced the first spectroscopic atlas of OB-type stars in Sextans A. From the whole sample of 18 observed stars, 12 were classified as early OB-types, including 5 O-stars. The radial velocities of all target stars are in agreement with their Sextans A membership, although three of them show significant deviations. We determined the stellar parameters of the O-type stars using the stellar atmosphere code FASTWIND, and revisited the sub-SMC temperature scale. Two of the O-stars are consistent with relatively strong winds and enhanced helium abundances, although results are not conclusive. We discuss the position of the OB stars in the HRD. Initial stellar masses run from slightly below 20 up to 40 solar masses.}
 {The target selection method worked well for Sextans A, confirming the procedure developed in Garcia \& Herrero (2013). The stellar temperatures are consistent with findings in other galaxies. Some of the targets deserve follow-up spectroscopy because of indications of a runaway nature, an enhanced helium abundance or a relatively strong wind. We observe a correlation between HI and OB associations similar to the irregular galaxy IC1613, confirming the previous result that the most recent star formation of Sextans A is currently on-going near the rim of the H\,{\sc I} cavity. }


        \authorrunning{I. Camacho et al. }
        \titlerunning{New OB-stars in Sextans A}

   \maketitle
%

\section{Introduction}                   

Massive stars are primary engines of the dynamical and chemical evolution of the Universe.
Their strong radiation-driven winds, ionizing radiation fields and violent death strongly affect the evolution of the galaxies and the Universe. 
 
Star formation models indicate that first generation stars were really ma\-ssi\-ve  \citep[$M_{*} \gtrsim 100$ \Msun,][]{Bromm02}, and may have started the re-ionization of the Universe \citep{Robert10}.
It is important to understand the pro\-per\-ties of \-ma\-ssi\-ve stars, both fundamental parameters and evolution, in nearby galaxies with as similar chemical composition as possible to that expected in the early Universe. Very low-metallicity stars are particularly interesting in this context, as the primitive Universe \citep[and the first very massive stars,][]{BFC01} had roughly zero metallicity.

Some dwarf irregular galaxies of the Local Group beyond the Small Magellanic Cloud (SMC) represent a significant decrease in metallicity  while still enabling studies of resolved populations. Their stars are not as massive as Population III ones, but their low metallicities and high temperatures make them the next logical step in the exploration of the physics of the first population of stars.

 Several projects over the past few years had as primary objective to cha\-rac\-te\-ri\-ze the relationship between the physical properties of ma\-ssi\-ve stars and the initial metal composition of the gas where they formed. Studies like those performed by \citet{K02}, \citet{Evans03}, \citet{MML04}, \citet{MPP05}, \citet{MKE06}, \citet{Trundle07}, \citet{Tramper11,Tramper14}, \citet{GH13} (hereafter [GH13]) and \citet{GH14} analyzed the dependence of effective temperature, wind properties or ionizing flux on metallicity.

Metallicity also alters the mechanism responsible for the mass loss of massive stars, driven by radiation pressure. Theory predicts a strong correlation between the momentum ca\-rried by the stellar wind and the luminosity and metallicity of the star: the Wind-Momentum Luminosity Relationship (WLR, \citet{Kal95}). This relationship between the radiative wind and the metallicity of the star has been already characterized theo\-re\-ti\-ca\-lly  \citep{Vink01}, predicting a decrease of the wind mass-loss rate (\Mdot) and momentum with decreasing metallicity. This has also been confirmed observationally  
 by spectroscopic studies of massive stars in the Milky Way (MW), Large Magellanic Cloud (LMC) and SMC, covering the me\-ta\-lli\-ci\-ty range from 1 to 0.2 \Zsun $\,$ \citep{MKE07b}.

Confirmation of the validity of the WLR at lower me\-ta\-lli\-ci\-ties  is a crucial  step towards the physics of the early Universe. However, there have been some indications of a possible breakdown of this relationship  at $Z<Z_{SMC}$. Herrero et al. (2010,2012) analyzed an O7.5(f) III star in IC1613 ($ Z = 1/7$\Zsun $\,$ from O abundance determinations) finding a wind momentum larger than predicted by theory. 
Tramper et al. (2011, 2014) analyzed a more extended sample of 10 stars in IC1613, WLM and NGC3109 and found again wind momenta larger than predicted by theory. These works are based solely on optical spectroscopy and suffer from uncertain terminal velocities estimated from the escape velocity, affecting the WLR \citep{Herrero12,GH14}.

Moreover, the independent works of \citet{GH14} and \citet{Hosek14} have found evidence that the Fe abundance in IC1613 and NGC3109 may be higher than the 1/7Fe$_{\odot}$~value inferred from oxygen. As Fe is the main driver of the mass-loss, this mismatch could explain the strong winds problem. The result for IC1613 has been confirmed by \citet{Bouret15}, who also find  SMC-like iron abundances. The WLR remains to be tested in galaxies of poor iron content. \\

Sextans A (DDO 75, $\alpha$:10:11:00.5 $\delta$:-04:41:29, J2000.0) is a metal-poor irregular galaxy in the outer part of the Local Group, and the analysis of a sample of young stars has revealed very low Fe abundance {$\langle (FeII, CrII)/H \rangle=-0.99 \pm 0.04 \pm 0.06$ \citet{KVT04}, where the first error bar is the standard deviation of the line-to-line measurements, and the second represents the systematic error in the element abundance from the uncertainties of the derived stellar parameters. In fact, Sextans A is the Local Group  galaxy where the most Fe-poor young stars have been found until date \citep[see Fig. 16,][]{Hosek14}. For this reason, Sextans A offers the unique opportunity to study massive stars in a really metal-poor environment. \\

This galaxy was being forming stars intensely 1-2 Gyr ago, and with a remarkable increase the last 0.6 Gyr \citep{Dolphin03b}. Three important regions of star formation are visible in Sextans A: The oldest is located in the north of the galaxy (region A, see Fig. \ref{F:targetsyextras}), with an approximated age of 400 Myr; another is in the east-south (B) forming stars for at least 200 Myr and the youngest one is situated in the north-west (C), with an age lower than 20 Myr \citep{Dohm02}. Focusing on the youngest generation, \citet{Bal12} found two populations of a few million years and a few tens of million years respectively. The presence of young hot massive stars in the galaxy is indeed hinted by a number of H\,{\sc II} bubbles and superbubbles in and around region-B, and more diffuse ionized structures around region-C (see Fig. \ref{F:targetsyextras}). \\

In spite of the presence of these star-forming regions, Sextans A has a low Star Forming Efficiency (SFE), as expected for very low metallicity environments \citep{Shi14}. This low SFE is associated to a large amount of molecular gas, that however does not result in more star formation. Shi et al. suggest that intense radiation fields may heat the molecular gas, inhibiting star formation. Massive stars are natural sources of intense and energetic radiation fields, particularly in low metallicity environments, where the UV opacity in their atmospheres is reduced. This situation may resemble that of the first epochs of star formation in the early Universe. \\

Considering its low metallicity, young population and proximity \citetext{\citealp[distance modulus $\mu_{0}=25.60 \pm 0.03$,][]{Tammann11}; \citealp[foreground reddening $E(B-V)=0.044$,][]{Skillman89}} , Sextans A is an excellent candidate for studying  blue massive stars in the most Fe-poor environment of the Local Group. 
The goal of this paper is to unveil new OB stars in the galaxy using low-resolution spectroscopy on its resolved population. The confirmed candidates will be subsequently followed-up with higher resolution optical and ultraviolet spectroscopy for a detailed study of their photospheric and wind parameters, to ultimately characterize radiation-driven winds at low metallicity. \\

This paper is structured as follows: In Sec. 2, we provide details of the candidate selection process, discussing the applied criteria. We also discuss the data reduction process and the instrumentation used. The spectral classification is presented in Sec. 3. In Sec. 4, we proceed to a preliminary spectroscopic analysis of the confirmed O-stars and the discussion of the results, including the evolutionary status of our sample stars. We also revisit the temperature scale for the sub-SMC metallicity regime proposed by [GH13]. Finally, in Sect. 5, we summarize our conclusions.

%

\section{Observations and data reduction}   
\label{S:obs}
\label{target_selection}

The only means to unequivocally unveil the elusive OB spectral types is low-resolution spectroscopy. The reason is that O and early-B stars have very similar optical colors and the degeneracy cannot be broken unless extremely accurate multi-band photometry is available \citetext{\citealp[e.g.][]{AHS98} \citealp[and][]{Maiz04}}. \\

We performed low-resolution, long-slit spectroscopy on Sextans A with the Optical System for Imaging and low-Resolution Integrated Spectroscopy (OSIRIS) at the 10 meter telescope Gran Telescopio Canarias (GTC), under proposal GTC59-12A (P.I. A. Herrero).

Given the relatively small angular size of Sextans A (5' in \-dia\-me\-ter), the OSIRIS detector ( 7' $\times$ 7' field of view) completely covers the galaxy, and we were able to observe a total of 18 candidate OB stars with six 1.2" wide slits. The resulting spectra taken with the R2000B grating cover the 4000-5500 \AA $\;$ range with a resolving power of R $\sim$ 1000. \\

\begin{table*}
\caption{Observing log.} 
\label{T:log}      
\centering 
  
\begin{tabular}{|c c c c c c c c|}        
\hline\hline               
  Slit      &  Stars                      & Date       & Seeing [$\arcsec$]    & Trans.& Moon  &  Airmass  &  Exposure time [s]\\ 
 \hline                                                                                 
 slit\_1    & OB121$^{*}$, OB122                & 03-23-2012 & 1.2        & Clear + dust  & Dark &  1.42   & 1698\\
                                                                                        
 slit\_2    & OB221, OB222$^{*}$               & 03-23-2012  & 1.1      &  Clear + dust  & Dark  &   1.22  & 2498\\
                                                                                        
 slit\_3    & OB321, OB323$^{*}$, OB324, OB326  & 03-23-2012  & 1.1      &  Clear + dust  & Dark & 1.21    & 2698\\
                                                                                        
 slit\_4    & OB421, OB422                & 03-25-2012  &  1.1      & Clear + dust  & Dark & 1.28   & 2698 \\
                                                                                        
 slit\_5    & OB521, OB523, OB524, OB525  & 03-25-2012  &  1.1      & Clear + dust  & Dark & 1.40   & 2698 \\
 
 slit\_6    & OB621, OB622, OB623, OB625 & 03-26-2012  &  1.1      & Spectroscopic & Dark & 1.27    & 2698\\
\hline 

\end{tabular}

\tablefoot{ \\
\tablefoottext{*}{Stars included in two different slits\\}
}
\end{table*}

We used the target selection criteria of [GH13], which is optimized towards the O- and early-B types. Starting from the photometric catalogue of \citet{MO07}, we selected stars with a reddening-free pseudo-color $Q= (U-B)-0.72(B-V)<-0.8$, assigning higher priority to those targets with detection in GALEX from a visual inspection of images. We note that the Q-index is only reddening free for a standard $R_v=3.1$ law. However, we estimate that the effect of a different extinction law is comparable to (or even lower than ) the one inflicted by the reddening differences within the galaxy.

We mainly targeted sources brighter than  $V=19.6$ that could be observed in $\sim$1 hour long Observing Blocks (OBs) with sufficient signal to noise ratio (SNR) for spectral classification. We note that this upper limit for the V-magnitude may have left out the reddened massive stars embedded in gas and dust. As a secondary criterion, visually bright targets coincident with strong GALEX detection were also included regardless their Q-index. 

This makes a total of 11 candidate stars.  6 of them were chosen because they met the main Q-value  criterion, and 4 due to their  strong GALEX emission. The eleventh candidate is \citet{KVT04}'s SexA-513 star, classified by the authors as  F-hypergiant, to check if the star has experienced changes in spectral type and establish a possible connection with the Luminous Blue Variable (LBV) stage. Since our slits cross the entire galaxy, they included 7 additional stars observed with enough SNR, out of which 4 have  $Q<-0.8$. Thus we observed a total of 18 stars, from which 10 fulfill our main selection criterion. \\

Details of the observations are provided in Tables \ref{T:log} and \ref{T:assoc}; on-target exposures range from 1700s to 2700s. Fig. \ref{F:targetsyextras} shows the distribution of the program stars and slits in the galaxy. \\

\begin{figure}
\centering
\includegraphics[width=0.45\textwidth]{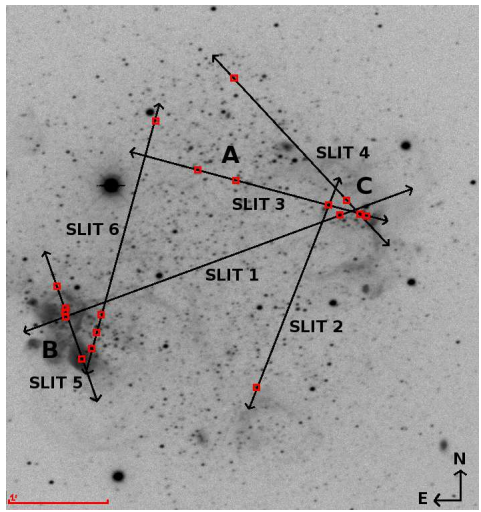}
   \caption{H$_{\alpha}$ image of Sextans~A  (from \citet{MO07}; available at http://www.lowell.edu/users/massey/lgsurvey.html). North is up and East to the left. The orientation of the chosen slits are shown together with the position of the candidate OB stars, marked with red squares. The 1 arcmin lengthscale is provided at the bottom of the Figure. The main star forming regions in Sextans A are marked (A-C).
           }
      \label{F:targetsyextras}
\end{figure}

 \begin{table*}
\caption{Program stars. Identification numbers, coordinates and photometry by \citet{MO07} are also provided. The spectral types (column SpT) and radial velocities were derived in this work. The radial velocities are provided in the heliocentric standard of rest. The signal to noise ratio (column SNR) is given per resolution element. }             
\label{T:assoc}      
\centering          
\begin{tabular}{l c c c c c c c c c}     
      
\hline\hline  
  ID & ID                  & RA(J2000.0)   & DEC(J2000.0)  &     V &   B-V  &     Q &          SpT  &  SNR   & $V_{rad}$                       \\
 This work & \citet{MO07}               & [h:m:s]  & [deg:m:s]         &       &        &       &            &      &  [\kms]            \\
\hline

OB326$^{C \dag \boxtimes}$ & J101053.81-044113.0     &    10:10:53.81 &       -04:41:13.0 & 20.688  & -0.265  & -1.015 &  O7.5~III((f)) &  35 & 331 $\pm$         20\\

OB623$^{B \dag }$ & J101104.78-044224.1     &    10:11:04.78  &     -04:42:24.1  & 20.681  & -0.234  &-1.001 &   O8~Ib &  30  & 330 $\pm$    40\\

OB521$^{B}$ & J101105.38-044240.1     &    10:11:05.38 &      -04:42:40.1  & 19.459  &  -0.259 & -0.950 &   O9.5~III-V &  55 & 343 $\pm$            30\\

OB523$^{B}$ & J101106.05-044211.4    &    10:11:06.05 &      -04:42:11.4  & 19.492	&  -0.231  & -1.013 &   O9.7~I((f)) & 54  & 442 $\pm$           30\\

OB321$^{A}$  & J101100.66-044044.3     &    10:11:00.66 &      -04:40:44.3  & 19.609	&  -0.248 & -0.960 &   O9.7~I((f)) &  47 & 327 $\pm$         20\\

OB622$^{A}$ & J101102.38-044014.6    &     10:11:02.38  &      -04:40:14.6 & 19.581	& -0.099  & -0.683 &   B0~I &  47  & 313 $\pm$         20\\

OB524$^{B \dag}$ &  J101106.03-044209.1     &    10:11:06.03 &      -04:42:09.1  & 20.834	&  -0.272 & -0.907 &   B0~III &  37   & 301 $\pm$           20\\

OB222$^{C}$ &  J101055.35-044106.1    &    10:10:55.35 &      -04:41:06.1  & 19.067	&  -0.171  & -1.007 &   B1~I &  90  & 247 $\pm$         20\\

OB421$^{A}$ & J101059.21-043948.1     &    10:10:59.21 &      -04:39:48.1  & 19.329	&  -0.192 & -0.880 &   B1~I & 32   & 378 $\pm$           40\\

OB422$^{C \dag}$ & J101054.62-044103.0     &    10:10:54.62 &      -04:41:03.0  & 19.739	&  -0.218 & -0.942 &   B1~I &  38  & 312 $\pm$           30\\

OB323$^{C}$ &  J101054.08-044111.5    &    10:10:54.08 &      -04:41:11.5  & 19.487  &  -0.217 & -0.875 &   B1~III &  78  & 310 $\pm$           20\\

OB221 &  J101058.27-044257.8    &    10:10:58.27 &      -04:42:57.8  & 19.176	&  -0.074 & -0.706 &   B2.5~I & 49  & 301 $\pm$           20\\

OB324$^{A \dag \boxtimes}$ & J101059.13-044051.0     &    10:10:59.13 &       -04:40:51.0 & 20.486  & -0.069  & -0.412 &   B8~II &  38 & 326 $\pm$ 20         \\

OB121$^{B}$ &  J101106.02-044214.2    &    10:11:06.02 &	 -04:42:14.2   & 18.771	&  0.054  & -0.277 &   B8~II & 47  & 335 $\pm$ 30           \\

OB122$^{C}$ & J101054.89-044112.1     &    10:10:45.89 &      -04:41:12.1  & 18.480	&  -0.010 & -0.529 &   B9~I &  54 & 322 $\pm$  30         \\

OB525$^{B \dag}$ &	J101106.38-044155.9     &    10:11:06.38 &      -04:41:55.9  & 20.519	&  -0.043 & -0.155 &   A0~II & 31  & 352 $\pm$           10\\    

OB625$^{B \dag}$ & J101104.58-044213.0     &    10:11:04.58  &     -04:42:13.0  & 20.668  &  0.267  &-0.274 &   A5~II &  22 & 302  $\pm$ 30\\

OB621$^{B \star}$ & J101104.97-044233.8     &    10:11:04.97 &      -04:42:33.8  & 17.477	&  0.174  & -0.188  &   F5~I & 57  & 320 $\pm$          10\\

\hline                  
\end{tabular}
\tablefoot{ \\
\tablefoottext{A,B,C}{The star is located in one of the three regions of star formation of Sextans A\\}
\tablefoottext{\dag}{Extra stars (not primary program stars) included in the slit \\}
\tablefoottext{\boxtimes}{Off-slit stars \\}
\tablefoottext{\star}{Star presented in  \citet{KVT04}\\}
}
\end{table*}


The reduction process was performed with standard IRAF\footnote{Image Reduction and Analysis Facility. IRAF is distributed by the National Optical Astronomy Observatory, which is operated by the Association of Universities for Research in Astronomy (AURA) under cooperative agreement with the National Science Foundation.} procedures. For a more detailed description, we refer the reader to [GH13]. The images were trimmed to remove vignetted areas of the CCD with the \textit{ccdproc} routine, also used to apply the overscan and flat-field corrections. The spectral extraction and background subtraction was performed using  \textit{apall}. The keywords for optimal extraction were switched off, after checking they introduce artifacts on our low SNR data. The two half exposures of each OB, split to enable cosmic ray removal, were coadded before the \textit{apall} extraction.  Finally, the extracted spectra were normalized by fitting the continuum to a smooth function, using the \textit{continuum} routine. \\

We note that a ghost effect, affecting wavelengths between 4743-4746 $\AA$, persists after the reduction process although this has no impact in our analysis. We also note that several targets are located in highly populated regions and experience nebular contamination, which hindered the background extraction. \\

We included targets OB121, OB222 and OB323 in two slits to check for internal consistency. After the positive check, the resulting spectra were coadded for increased SNR, their total exposure time being 4396, 5196 and 5396 s respectively.

\begin{figure*}
\centering
\includegraphics[width=\textwidth]{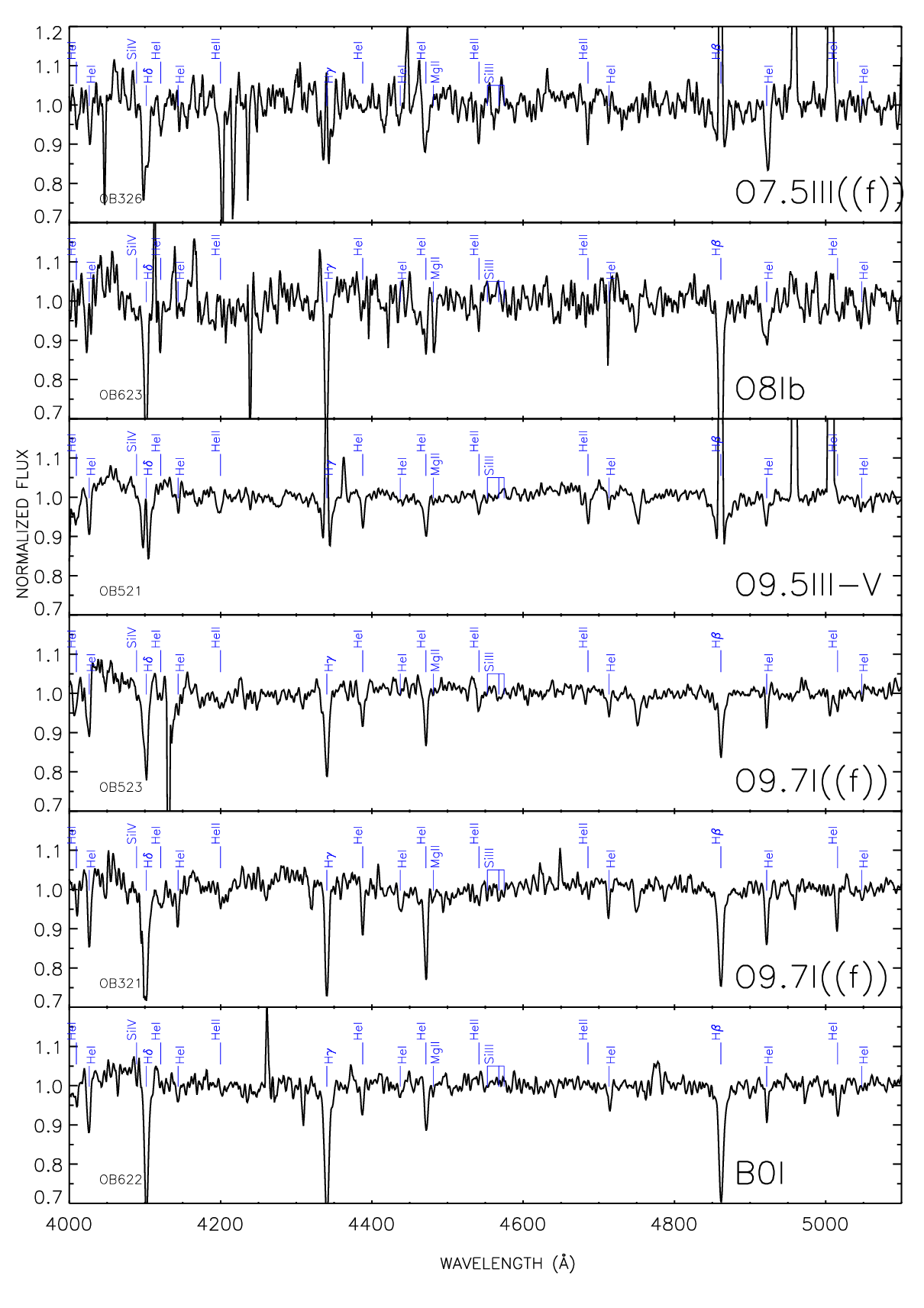}
   \caption{GTC-OSIRIS spectra of the program stars. The spectra have been smoothed for clarity, resulting in an effective resolution of about 800 for the plotted spectra. The spectra have been also corrected by the stellar radial velocities.
           }
      \label{F:spec1}
\end{figure*}

\begin{figure*}
\centering
\includegraphics[width=\textwidth]{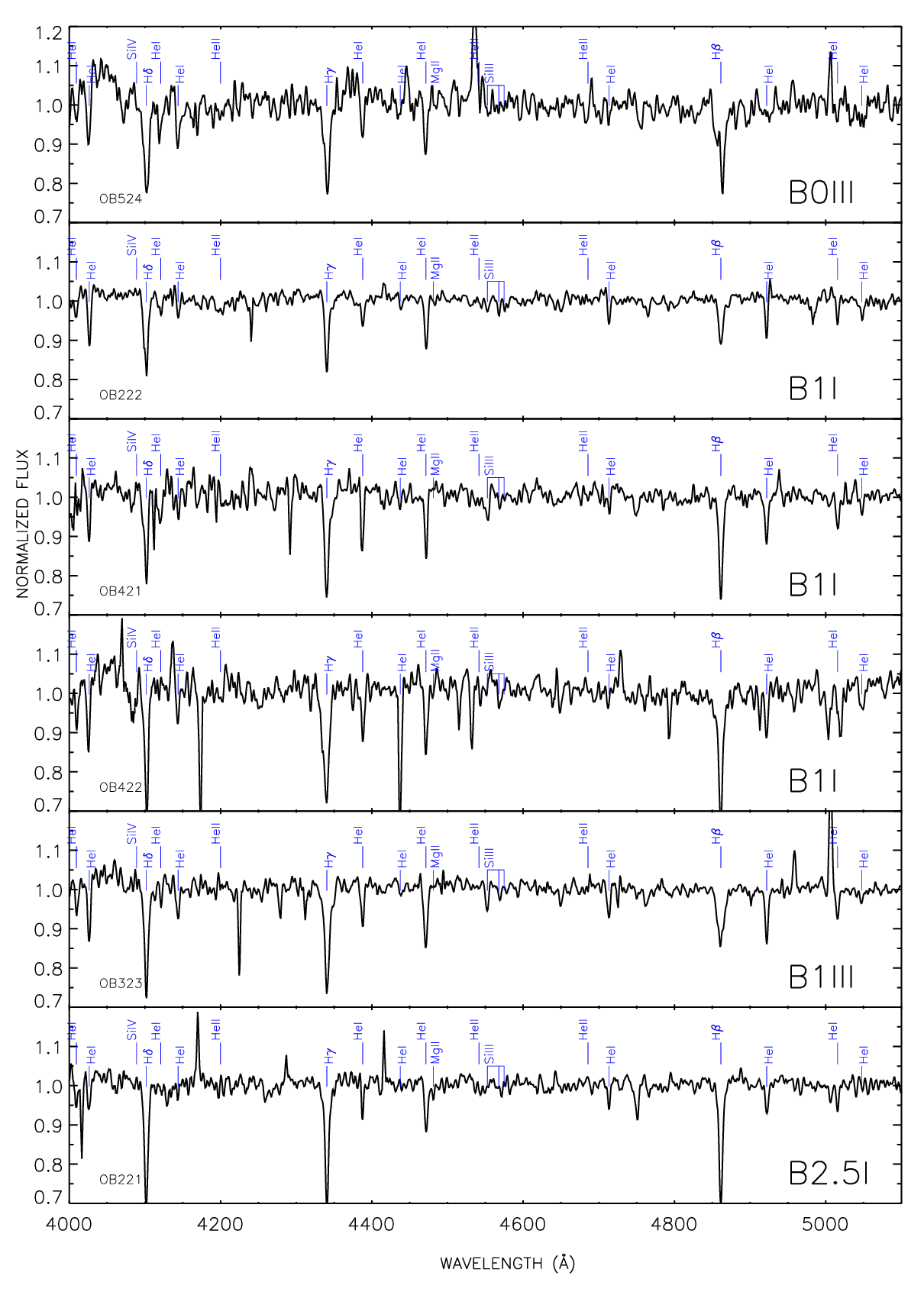}
   \caption{continued.
           }
      \label{F:spec2}
\end{figure*}

\begin{figure*}
\centering
\includegraphics[width=\textwidth]{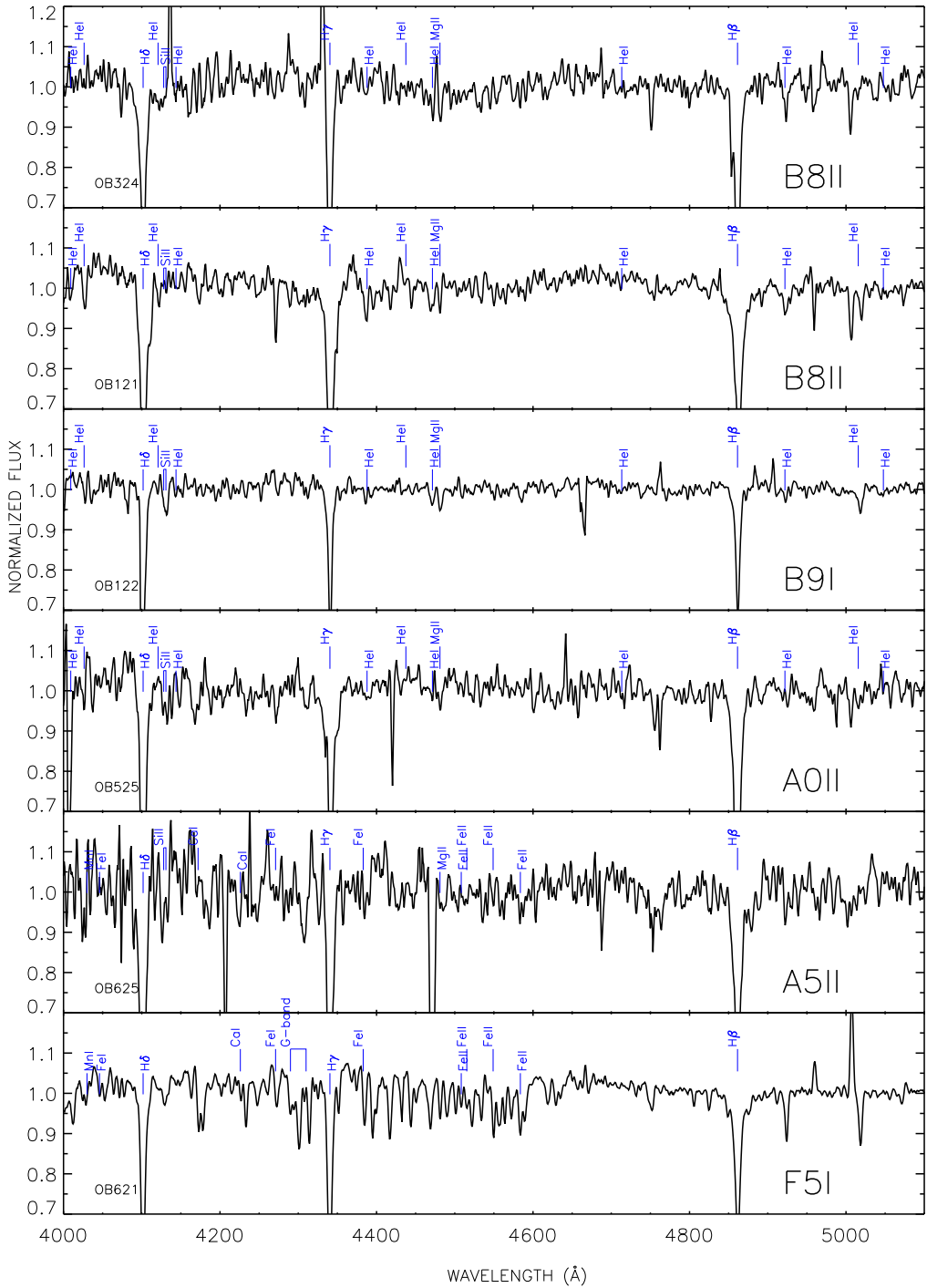}
   \caption{continued.
           }
      \label{F:spec3}
\end{figure*}

The reduced stellar spectra are shown in Figs.\ref{F:spec1}, \ref{F:spec2} and \ref{F:spec3}. The internal target identification code follows these rules: the first two letters, OB, indicate that they are stars belonging to one of the six Observing Blocks. The first number indicates the slit, the second one is the CCD chip, and the last digit is the number of the star in the particular slit. In  table \ref{T:assoc} we also provide the ID in the Massey et al. catalogue, allowing an unambiguous identification.


\section{Radial velocities and spectroscopic classification}        
\label{S:analisis}

\subsection{Radial velocity determination}    

We estimated the radial velocity of each star from the Doppler shift of a number of selected spectral lines, whose centroid was determined by fitting the profile to a gaussian function. For O- and B-type stars we used the lines of He\,{\sc I(+II)} 4026, He\,{\sc I} 4387 and He\,{\sc I} 4471 together with Mg\,{\sc II} 4481 for intermediate and late B stars. For later-type stars, we used the Balmer series in order to ensure the correct detection of the lines. The values derived for individual lines were averaged, and transformed to the heliocentric restframe. 
The final radial velocities, shown in Table \ref{T:assoc}, have a characteristic error of 20-30 \kms. \\

In Fig. \ref{F:vels} we compare our results with the radial velocity curve of Sextans A determined from radio observations of H\,{\sc I} \citep{Skillman88}. The black solid line represents the velocity curve, and the gray shaded areas delimit the $\pm$ 30 \kms \, uncertainty of their results. \\

Most  target stars present velocities within the expected range of values, in agreement with Skillman results. However, OB222, OB421 and OB523 present significant deviations.  We checked that their radial velocity measurements are not affected by the presence of nebular emission. These targets are well included in the galaxy and do not cluster in a particular location, lying in the C, A and B forming regions, respectively. Since their spectral classification and visual magnitude are consistent with the stars belonging to Sextans A, the possible explanation is that the stars are runaways or that they belong to a binary system. Confirmation of these hypotheses requires spectroscopic follow up.

\begin{figure}
\centering
\includegraphics[angle=+90,width=0.50\textwidth]{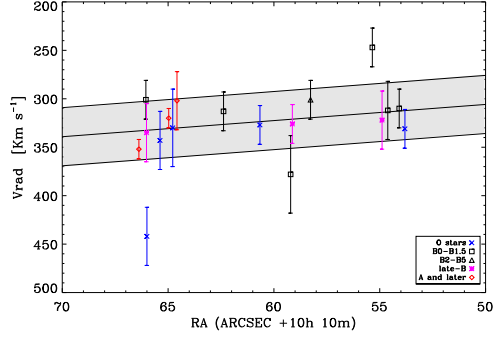}
   \caption{Radial velocity curve of Sextans A. Symbols mark the heliocentric velocities obtained for our  sample stars,with spectral types coded as indicated in the legend. The thick solid line represents  the mean value of the radial velocity obtained by \citet{Skillman88}. The shaded areas mark their $\pm$ 30 \kms \, uncertainty.
           }
      \label{F:vels}
\end{figure}

\subsection{Spectral classification}          

The spectral classification was based on \citet{Castro08}'s scheme for early massive stars, that uses He and Si diagnostic lines adapted for low metallicity environments.
We complemented these with the criteria proposed by  \citet{Sota11} for late O-type stars. The latter were established for Galactic stars, but are still suitable for metal-poor regions if only criteria involving He lines are used.

 We used criteria involving only He or only metal lines whenever possible. The use of mixed  criteria generated an uncertainty in the derived spectral classification that  depends on the spectral type, and it is more acute around $\sim$O9 types. For the stars classified with helium or metal lines separately, the uncertainty is expected to be smaller. We estimate the uncertainty to be two spectral subtypes in the first case and one in the second case.

Spectral sub-types for O-stars were derived by com\-pa\-ring mainly the He\,{\sc II} 4541/He\,{\sc I} 4471 and He\,{\sc II} 4200/He\,{\sc I(+II)} 4026 ratios. For late O-types  we also used the relative strength of He\,{\sc II} 4541/He\,{\sc I} 4387 and He\,{\sc II} 4200/He\,{\sc I} 4144.  The luminosity class of O-type stars was assigned according to the intensity of the He\,{\sc II} 4686 line and whether it was present in emission or absorption.

The main diagnostics for the earliest-B spectral sub-types were the pre\-sen\-ce of Si transitions in different ionization stages, such as Si\,{\sc IV} 4089 and Si\,{\sc III} 4552. We evaluated the relative strength of these lines to differentiate between contiguous subtypes. The blends of O\,{\sc II}+C\,{\sc III} at 4007 and 4650 \AA $\;$ also helped us to distinguish between subtypes B1 and B2, although with a lower weight because of the limited SNR. The relative strength of He\,{\sc I} 4471/Mg\,{\sc II} 4481 was also used as indicator from B1 until the latest B subtypes. Finally, the luminosity class was determined from the width of the Balmer lines.

Our  observations do not cover the main criterion to classify low-metallicity A stars (the Ca\,{\sc II} doublet at 3933-3968 \AA). Instead, we used the relative strength of Mg\,{\sc II} 4481 to the Fe\,{\sc II} 4508,4515,4549 and 4584 lines as criteria. Luminosity class was mainly assigned attending to Balmer line width.
Later spectral types were classified after \citet{gray09}. \\

The star OB621 was previously classified as an F-hypergiant by \citet{KVT04}. However, both our spectrum and the derived absolute magnitude (see Sec. \ref{S:HRD}) indicate a normal supergiant. Thus, we confirm the spectral type from Kaufer et al., but not the luminosity class.  \\

The resulting spectral types are provided in Table \ref{T:assoc}. Comments on individual targets can be found in the on-line Appendix-A. \\

\subsection{Success rate}         

\label{SS:success}
 
Even at low resolution, the observing time needed to obtain stellar spectra in nearby galaxies is significant. Therefore, an assessment of the criteria used for the candidate selection is relevant. 
\citet{GHV09} showed that OB-stars in IC1613  are found in a particular locus of the U-B vs. Q diagram (Box 1, see Fig. \ref{F:QUB_SextansA}). To further favor the discovery of O- over B-type stars [GH13] set a tighter constrain on Q  (Q$<$-0.8).\\

 Out of 10 sample stars with Q$<$-0.8, 5 are O stars and all of them have spectral type B1 or earlier. While the observations were satisfactory the number of unveiled O stars is small, and  we find 4 B1 stars in the group of Q$<$-0.8.  This makes a success rate of 50\% for O-stars, in contrast with the 70\% rate of Garcia and collaborators. \\

 The selection criterion based  on strong GALEX emission without complying with the Q main criterion, has proved not to be efficient in the search for O-stars. The 4 stars  selected  based only on this criterion have turned out to be late supergiant-B stars. A possible explanation could be that the strong GALEX emission we observe proceeds from nearby, hotter, early-type stars.\\

 In Fig. \ref{F:QUB_SextansA} we compare the location of our Sextans A program stars in the U-B vs Q diagram with the position of IC1613's OB stars. We find that the O stars of Sextans A are systematically located towards bluer U-B colors. This shift could be caused by a different amount of foreground reddening towards the galaxies (but note that  $ E(B-V)_{IC1613}$, with a value of 0.02 from \citet{Lee93}, is smaller than Sextans A's) or a varying extinction law. Variations of the internal reddening will increase the scatter of the sample in both galaxies. But also, the alleged poorer iron content of Sextans A could make the stars slightly hotter (hence bluer colors) than IC1613 analogs with the same spectral types. In sight of Fig. \ref{F:QUB_SextansA}, we concluded that for an optimized search of O-stars, the criteria have to be adapted for each individual galaxy.

\begin{figure}
\centering
\includegraphics[width=0.50\textwidth]{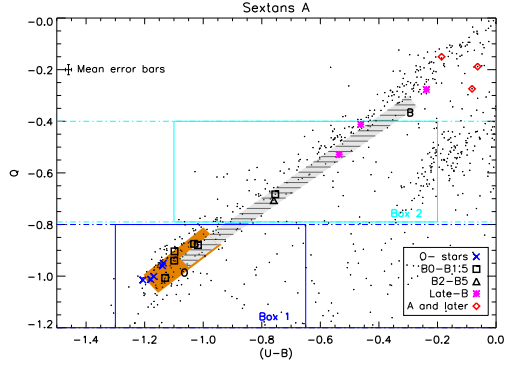}
   \caption{Sextans A U-B vs Q diagram. The black dots mark the photometric catalog of \citet{MO07}. Other colors and symbols represent the spectral classifications assigned in this work, as listed in the legend. The mean error bars for the program stars are shown in the upper left corner. The shaded boxes mark  the location of IC1613's O (orange) and B (gray filled with parallel lines) stars in the diagram (see [GH13]). The O-stars of Sextans A are offset towards bluer colors from the box of IC1613 O-stars.
           }
      \label{F:QUB_SextansA}
\end{figure}


\section{Analyses and discussion}            

In this section we discuss the properties of the sample stars as a population. As a necessary previous step, we determined the stellar parameters of the subset of O-stars.

\subsection{Spectroscopic analysis}                       
\label{S:analysis}

We analyzed the sample of O-stars with the automatic fitting program IACOB-GBAT \citetext{\citealp[described in][]{Simon11} \citealp[and][]{Carol14}}. IACOB-GBAT compares the observed spectra with an extensive grid of synthetic spectra, and finds the best fitting model using a $\chi^2$~algorithm. We used the Z=0.13\Zsun~grid developed by [GH13], since the impact of the small metallicity difference with Sextans A (0.1 \Zsun) is negligible compared to other uncertainties involved in the analysis. The grid covers effective temperatures from 25000 to 55000 K in 500~K steps, gravity runs from \logg=2.6 to 4.3 and the wind-strength Q parameter ($  Q= \dot{M} /(v_{\infty}R_{*})^{1.5}$) from -12.0 to -15.0. The models were calculated with FASTWIND \citep{FW97,FW05} accounting for line blanketing and NLTE effects in a spherically-symmetric atmosphere, and unclumped stationary radiation-driven winds. \\

The low SNR and low spectral resolution of our dataset limit our ability to determine all stellar parameters at once. We have therefore fixed the helium abundances ($\epsilon_{He}$), the exponent of the wind velocity law ($\beta$) and the microturbulence ($\xi$) to values typical for this kind of stars: $\epsilon_{He}$=0.09,$\,$ $\beta$=0.8 \citep{Kudri00} and $\xi$=10 \kms \citep{MKE06}. We also used a typical constant value for the projected rotation velocity of metal-poor O-stars of vsini=80 \kms.  This value has been adopted from \citet{RAal2013} for the LMC. Because of the good agreement between the LMC distribution and the results on Galactic stars by \citet{Simon14}, we do not expect large metallicity effects on the stellar rotational velocities.\\

In the view of the heavy nebular contamination of the Balmer lines in most of the sample stars, which may also affect He\,{\sc II} 4686, we consider that we do not have enough constraints to accurately determine the wind Q-parameter either. Because of the low metallicity of Sextans A we do not expect strong winds in our stars, although we should bear in mind the possibility of winds stronger than predicted by the theory of radiatively driven winds at very low metallicities (see the discussion in the Introduction). Hence, we carried out extra tests to check whether a different value of $\epsilon_{He}$~and logQ could improve the fit (see below).\\

The physical parameters derived for the sample of O-stars are provided in Table \ref{T:FASTWIND}. The large reported uncertainties stem from the low spectral SNR. Yet, the derived effective temperatures match what is expected from the spectral types (see below), and the two stars classified as giants/dwarfs have larger gravities than the supergiants. Considering the spectral quality the fits, shown in Fig. \ref{F:fits}, are also good.
We set the IACOB-GBAT code to fit only the strongest spectral features: $H_\gamma$, $H_\beta$, He\,{\sc II} 4686, He\,{\sc II} 4541 and He\,{\sc I} 4471. \\

We did not attempt to fit the core of the Balmer lines, masked by the nebular contamination, but their wings. This is clearly seen in Fig. \ref{F:fits}. The nebular lines have been undersubtracted from the spectra of OB326, OB521 and OB523; the core of the synthetic spectra calculated for their Balmer lines is deeper than the observations, but the wings are well fitted. Conversely, the nebular lines were oversubtracted from the spectrum of OB623, and the observed Balmer lines are deeper than the model. \\

We also checked whether a different He abundance could improve the fit to the observed spectra. Two stars, OB326 and OB523, do show evidence of an increased He abundance (without a clear improvement in the accuracy of the parameters).  Their spectra are also better fitted with a stronger wind (log Q$\sim$ -12.5 dex). 
Interestingly, we also note that OB523 is one of the possible runaway stars in Fig. \ref{F:vels}. Confirmation of the enhanced helium abundances and winds would require a future, higher resolution spectroscopic follow up beyond the scope of this work.\\

The effective temperatures derived for the sample stars are consistent with the very-low metallicity \Teff-scale. [GH13] presented the first sub-SMC temperature scale for O and early-B stars, built from their own results on IC1613 (1/7\Zsun), plus five stars from \citet{Tramper11} and \citet{Herrero12}. We have updated the sub-SMC \Teff-scale with \citet{Tramper14}'s results on IC1613, WLM and NGC3109, and our results on Sextans A (see Fig. \ref{F:Teff_scale}).

The location of the O-stars of Sextans A is consistent with the sub-SMC temperature scale, but we are not able to improve the calibration. This was expected in view of the large uncertainties of our results, but it is also due to the large dispersion seen in the results for different stars. Besides the difficulty of the spectroscopic analysis at these large distances, this also  reflects that the spectral type-effective temperature calibration is not only a function of metallicity, but also of the evolutionary status and history of the stars, as recently pointed out by \citet{Simon14}. Nonetheless, the results indicate that the spectral classification is consistent with the automatic parameter determination.

\begin{table}[!h]
\caption{Parameters derived for O-type stars.}           
\centering        
\begin{tabular}{c |c c c }     
\hline\hline  
ID      & SpT                        & $T_{eff}$            & log g      \\
        &                            &    [kK]              & [dex]      \\
\hline
 OB326$^{\dag}$  & O7.5 III((f))     & 37.4 $\pm$ 7.9      &  3.80 $\pm$  0.50 \\
 OB623           & O8 Ib             & 31.4 $\pm$ 7.9      &  3.24 $\pm$  0.64 \\
 OB521           & O9.5 III-V        & 31.9 $\pm$ 3.8      &  3.72 $\pm$  0.37 \\ 
 OB523$^{\dag}$  & O9.7 I(f))        & 29.3 $\pm$ 4.4      &  3.27 $\pm$  0.55 \\ 
 OB321           & O9.7 I((f))       & 26.3 $\pm$ 4.3      &  2.90 $\pm$  0.30 \\
\hline
\end{tabular}
\label{T:FASTWIND}
\tablefoot{ \\
\tablefoottext{\dag}{Indications of strong winds\\}
}

\end{table}

\begin{figure}
\centering
\includegraphics[width=0.5\textwidth]{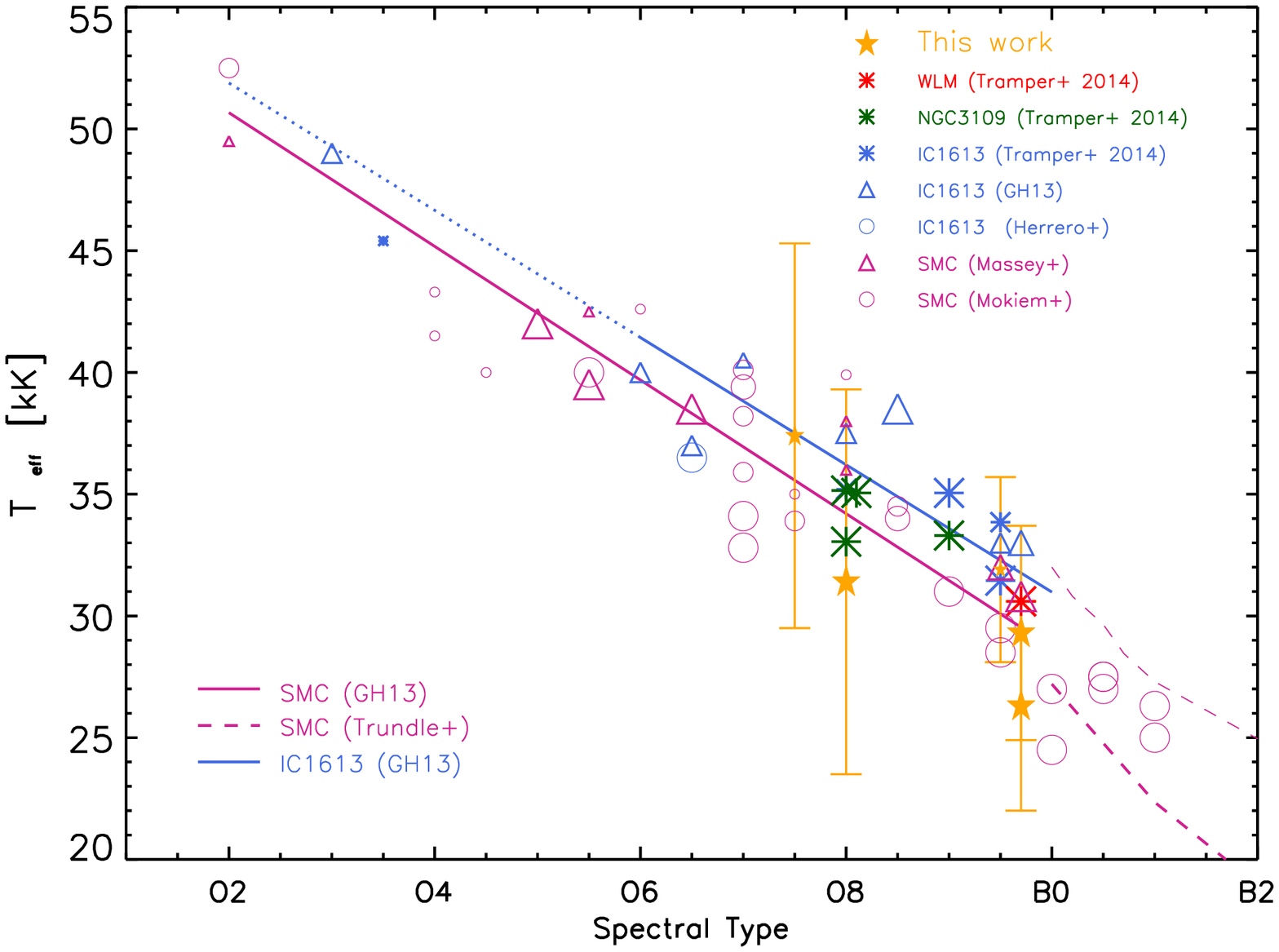}
   \caption{The very-low metallicity \Teff-scale, revisited. Symbols represent the effective temperatures of individual stars in IC1613, WLM, NGC3109 from [GH13], \citet{Tramper14} and \citet{Herrero12}, and in the SMC from \cite{MKE07a} and \citet{MZM09}. The host galaxy is  color-coded as indicated in the legend, whereas the symbol size codes the luminosity class (smallest symbol= dwarfs, largest symbol=supergiants). The solid lines are the \Teff-scales derived by [GH13] for IC1613 (blue) and the SMC (purple), and the dashed purple ones represent calibrations for SMC  B supergiants (dashed-thick) and dwarfs (dashed-thin). The position of NGC3109-4 in the abscissa axis has been slightly shifted to avoid overlap with NGC3109-1.} 
  
      \label{F:Teff_scale}
\end{figure}

\begin{figure*}
\centering
\includegraphics[width=\textwidth]{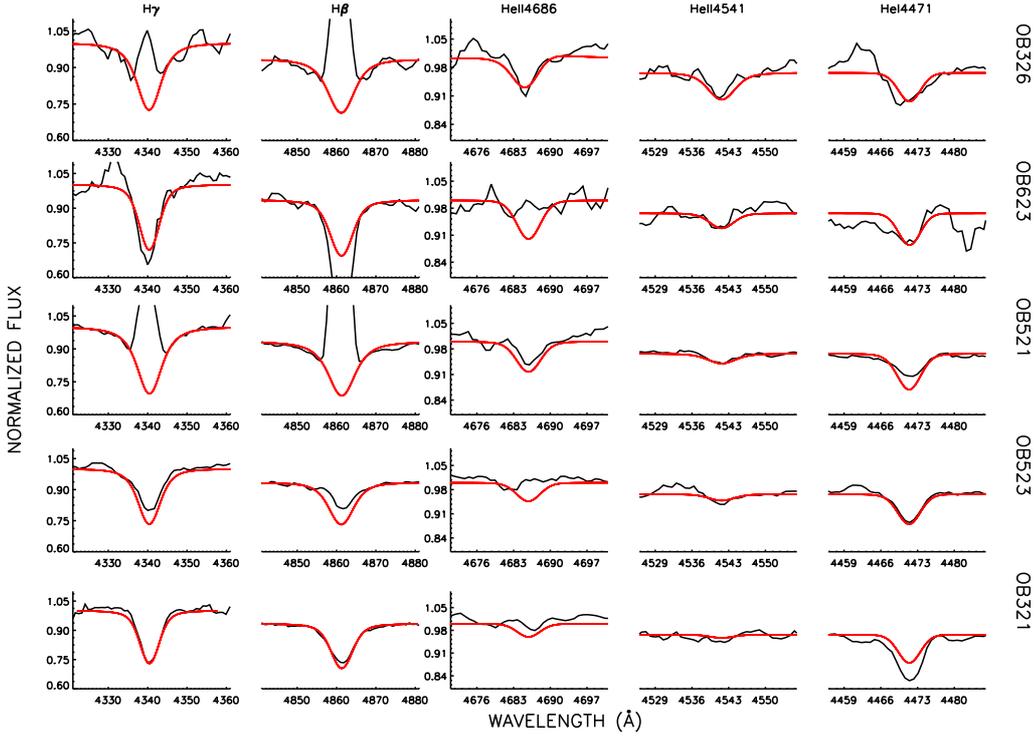}
   \caption{O-stars: OSIRIS observations (black) \textit{vs} a FASTWIND model calculated with the parameters derived in this work (red). The axes have been omitted in some of the plots for clarity sake. 
           }
      \label{F:fits}
\end{figure*}

\subsection{Hertzsprung-Russell diagram}      
\label{S:HRD}

The location of our program stars in the Hertzsprung-Russell diagram (HRD) is shown in Fig. \ref{F:HR}, together with tracks and isochrones for SMC metallicity from \citet{Brott11} without rotation.\\

\begin{figure*}
\centering
\includegraphics[width=\textwidth]{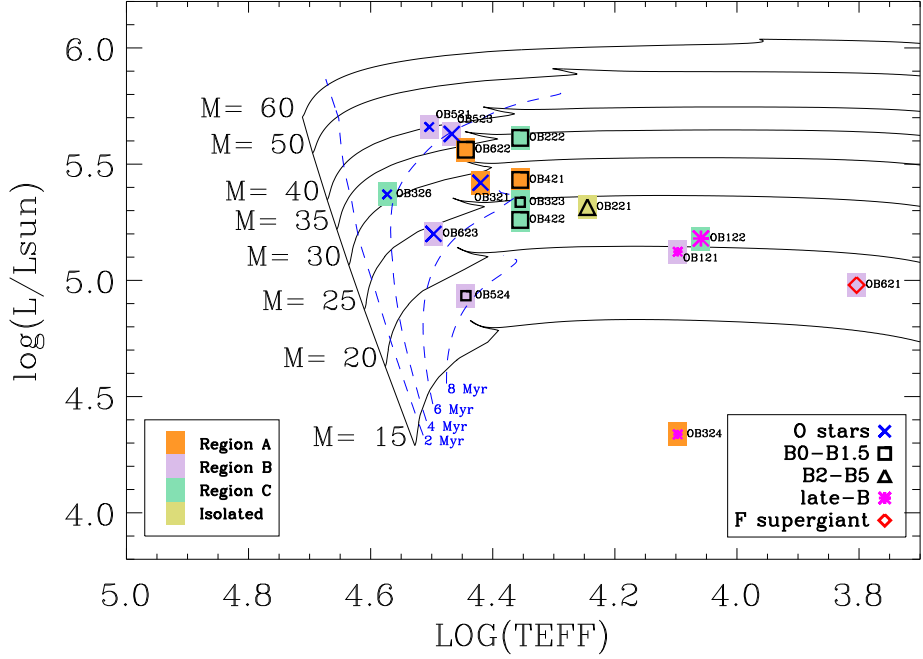}
   \caption{HR diagram for the OB stars analyzed in this paper. Different symbol size indicates different lu\-mi\-no\-si\-ty class (smallest symbol = dwarfs,  largest symbol = supergiants).
           }
      \label{F:HR}
\end{figure*}

To build the HRD, we used the effective temperatures derived in Sect. \ref{S:analysis} for the sample of O-stars. Lacking a more appropriate reference work, we used \citet{Markova08}'s \Teff~ calibration with spectral type for the B-type program stars, regardless their luminosity class. Although we note that this scale was defined from Galactic B supergiants and it may predict slightly cooler temperatures for our comparatively metal-poorer B-type sample stars, it is very similar to the results from \citet{Trundle04} and \citet{Trundle05} for the SMC and LMC \citep[see][Fig.10]{Markova08}. \\

Stellar luminosities for O stars were obtained from the absolute magnitudes, in the way explained in \citet{Herrero92}. For B stars, bolometric corrections from \citet{Balona94} were applied.

The absolute M$\rm_{V}$~magnitudes were calculated from the distance modulus  and \citet{MO07}'s photometry. To calculate the color excess we derived the individual $(B-V)_{0}$  colors with \citet{Massey00}'s prescription for 0.8\Zsun~  ($(B-V)_o = -0.005 + 0.317Q$). We preferred this over a calibration of $(B-V)_o$ based on spectral types because of the uncertainty of our classification.\\

For the F-supergiant star, the effective temperature and bolometric correction were taken from the calibration proposed by \citet{Allens00}.\\

The HRD shows that our sample O-stars are young and massive, with initial masses under 40\Msun~ and ages around 4-6 Myr. They all lie on the main sequence (MS). The most massive star, OB521, is at the center of a dense H\,{\sc II} bubble and possibly responsible for its ionization. The early-B supergiant stars, on the other hand, are already in post-MS positions, evolving towards a red-supergiant phase.

The uncertainties introduced by using tracks without rotation and the different metallicity between Sextans A and SMC are comparable. Combining  these two contributions, individual masses may be overestimated by 2-3 \Msun, and  ages by 1 Myr.  Had we considered a higher initial rotational velocity, the MS would slightly extend towards cooler temperatures, yet our conclusions on the evolutionary stage of the OB stars would remain the same. \\

The location of OB621 in Fig. \ref{F:HR} further supports our classification as an F-type supergiant. If the star was a hypergiant it would be located in the upper part of the HRD. Instead, it is among the least luminous of our sample stars. Given the 15-20\Msun~initial mass inferred from the HRD, it is unlikely that OB621 is an LBV-candidate. In addition, the star is well outside the usual locus of LBV-candidates \citep{SVdK04},  which have a lower limit of luminosity of $log(L/L_{sun})\sim$ 5.3. Besides being morphologically very different, the LBV-candidate we found in the metal-poor galaxy IC1613 is hotter and brighter \citep[\Teff=9260~K, log(L/\Lsun)=5.34,][]{Herrero10}.\\

Fig. \ref{F:HR} also allows us to detect OB324 (\Teff$\sim$12500~K, log(L/\Lsun)=4.34\Lsun), classified as B8~II as an outlier. Its blue color excludes a large extinction whose correction would result in a brighter absolute magnitude and would locate the star at the locus of B-supergiants. On the other hand, it is unlikely that OB324 is a background star: using the V-magnitude of the other B8~II sample star (OB121) as a reference, OB324 would be located at $\mu_o$=27.10. A possible explanation is that the star is actually a central star of a planetary nebula (CSPN) still in a low-excitation phase. 
We can estimate the absolute magnitude of a CSPN from  its theoretical luminosity and the bolometric correction. Neglecting metallicity effects, we use data from Allen's Astrophysical Quantities (2000, edited by Cox, chapter 15) and adopt a luminosity of log(L/Lsun)= 3.8 (a typical luminosity for CSPN, see \citet{Vassi94}). From the color-excess we estimate Av= 0.2. With these numbers we obtain an absolute magnitude of Mv= -4.5 for OB324. This corresponds to a distance of 1.1 Mpc, consistent with the known distance to Sex A.

\subsection{The latest star formation history of Sextans A}

The resolved populations of OB stars add enhanced time-resolution to the studies of the star formation history of galaxies at the most recent epochs. As we explained in the introduction, \citet{Dohm02} describe that star formation in Sextans A is concentrated in three regions.

Region-A at the North has been forming stars for the past 400~Myr but it is running out of gas and star formation will presumably not extend much longer. Region-B in the East-Southeast has been forming stars since 200~Myr ago and region-C in the West-Northwest was only activated 20~Myr ago. \citet{Dohm02} find indicators of star formation during the most recent time bin considered in their study (20 Myr) in the three regions, with the highest concentration of young stars in B and C. \\

We do not find any systematic age difference between the stars belonging to regions A, B or C, all in the $\sim$4 to 8~Myr range. Despite the fact that region-A is less compact and exhausting its gas reservoir, it still contains young and massive stars. These findings confirm that star formation has continued in the three regions until roughly the present epoch.  \\

We detect no special concentration of more massive stars in any particular region, thus no correlation of the most massive star with the column density of H\,{\sc I} (see below).

\citet{vDPW98} and \citet{Dohm02} found a strong correlation of their youngest sample stars with the distribution of neutral hydrogen in Sextans~A. The galaxy's visually brightest part is enclosed in a giant H\,{\sc I} depression, even though the cited works disagree on whether the cavity was carved by a giant supernova that exploded about 50 Myr ago at the galactic center and expanded outwards, or the cavity existed prior to the SN or the gas has been consumed by star formation. Although roughly round the cavity is not totally symmetric and the rim displays two bean-shaped over-densities on opposite sides \citep[see Fig. 14,][]{Dohm02}. The semi-major axis of the H\,{\sc I} distribution links these features, and matches the central concentration of RGB stars \citep{Dohm02} and the projected axis of a possible bar inferred from the H\,{\sc I} dynamics \citep{Skillman88}. \\

Both \citet{vDPW98} and \citet{Dohm02} conclude that the most recent star formation of Sextans A is currently on-going near the edge of the H\,{\sc I} cavity. In fact, regions B and C roughly match the H\,{\sc I} 'beans', while region-A and our sample star OB221 would be located in less dense regions of the rim. However, \citet{Dohm02} emphasize that while in region-C star formation is confined to the inner edge of the H\,{\sc I} over-density it fully overlaps with the central highest column-density of H\,{\sc I} in region-B, also the highest of the galaxy.

Our sample OB-stars consequently follow this distribution: region-C stars lie on the inner edge of the 'bean', but region-B stars are located in the 'bean'.

We found a similar correlation between H\,{\sc I} and OB associations in the dwarf irregular galaxy IC1613 \citep{GH10}. We detected OB associations in apparently inert regions of the galaxy that the inspection of radio data revealed to be on a H\,{\sc I} ridge. But the youngest OB associations of IC1613 agglomerate in the North-East part of the galaxy, that matches the highest density of neutral hydrogen and a number of spectacular H\,{\sc II} bubbles. The associations in the bubble region also present the largest age spread of the galaxy, implying that recent star formation had proceeded over a longer period of time than in the rest of the galaxy, and is still ongoing.  The most massive stars of IC1613 are also found in this region. 

Sextans A's densest H\,{\sc II} regions and shells also overlap with the highest density of H\,{\sc I}, i.e., region-B. Similarly to IC1613 our region-B sample stars show a variety of ages, and more young stars have been found in region-B than in C. This could seem at odds with \citet{Dohm02}'s statement that region-C is younger, but it rather indicates that even though C was activated later, B has sustained  vigorous star formation until recently. Qualitatively speaking this is confirmed by the H\,{\sc II} shell surrounding region-C, more diffuse and extended that the H\,{\sc II} structures around region-B. Since we can discard a lack of hydrogen in C, the expansion of this shell probably began earlier and no other mechanism could form a new bubble at later epochs. \\

\citet{Dohm02} concluded that the two sides of the galaxy containing regions B and C respectively have experienced the same spatially \textit{averaged} star formation history.  As an explanation, they suggest some global-scale regulatory process like the orbit with Sextans~B or perhaps a barred potential. Our findings suggest that the symmetry breaks in the last 10 Myr, although this period of time is smaller than the time resolution of \citet{Dohm02}'s work, 50 Myr. Another possible conclusion is that the regulatory agent works on timescales longer than 10 Myr.

Contrary to our results in IC1613, the stars we have found in region-B are not significantly more massive than the ones in A and C. However, recent Herschel and Spitzer observations have unveiled vast masses of dust in region-B \citep{Shi14} which are likely obscuring the youngest and most massive population. A deeper multi-object spectroscopic program to reach fainter O-stars is underway.


\section{Summary and future work}        
\label{S:con}

Low-metallicity galaxies offer an excellent proxy to explore  the physics of massive stars under conditions close to those of the early Universe. The nearby galaxy Sextans A has sub-SMC metallicity, and is thus an ideal laboratory to test our theories of stellar evolution and radiatively driven winds at those very low metallicities. \\

We have started a programme to observe and analyze massive OB stars in this galaxy. As a first step, we have selected candidate stars from photometric  properties for low-resolution spectroscopy and spectral classification, with the aim of future spectroscopy at higher resolution and SNR. \\

The selection criteria were adopted from the IC1613 study by [GH13]. The main criterion was to select V $<$ 19.6 stars detected in GALEX images with a reddening-free pseudo-color Q  $<$ -0.8 and following the position of OB stars in the Q vs (U-B) diagram. As a secondary criterion, we selected visually bright stars coincident with strong emission in GALEX images, even if they did not fulfill the main criterion.  While the main criterion may be affected by the assumed reddening law, we find this  has little impact on our main goal: to build a sample of O and early-B stars. The bright limiting  V magnitude introduces a bias against reddened stars, possibly the earliest O-types. Thus, the earliest spectral type that we find is O7.5 III ((f)). Finally, we have also included the star OB621, which was previously observed by \citet[][ SexA-513 for these authors]{KVT04}. \\

The observations were carried out with OSIRIS@GTC. They were successful in targeting blue massive stars and have produced the first atlas of OB-type  stars in Sextans A. Out of 17 candidates, we found five late-O stars (between O7.5 and O9.7) plus seven early-B stars (between B0 and B2.5). All 10 stars fulfilling the main Q-parameter selection criterion have spectral types B1 or earlier. Except for OB621 (that we reclassify  as a normal F-supergiant) all remaining stars are late-B or early-A supergiants. \\

Radial velocities are consistent with Sextans A membership for all targets, although three of them (OB222, OB421 and OB523) show significant deviations that may be caused by binarity or runaway status. \\

The quality of the spectra allowed us a preliminary spectroscopic analysis of the O stars in Sextans A. The parameter determination was limited to temperature and gravity, keeping all other stellar parameters fixed on a grid with Z = 0.13 \Zsun, previously developed by [GH13]. The results are consistent with the spectral classification and with the sub-SMC temperature scale by [GH13] and \citet{Tramper14}, although we are not able to improve the scale because of the limited accuracy of our results. Two of the stars (OB326 and OB523) showed indications of an increased He abundance and stronger winds in tests carried out by varying the stellar parameters. Interestingly, OB523 is one of the targets showing also radial velocity peculiarities.  \\

The position of the stars in the Hertzsprung-Russell Diagram is consistent with the expectations. The initial masses we obtain run approximately from 20 to 40 solar masses with ages concentrating around 4-6 Myr as compared to non-rotating SMC tracks by \citet{Brott11}. We estimate the effect of a different initial rotational velocity and a slightly different metallicity for Sextans A to be of the order of 2-3 solar masses and 1 Myr. Our findings are consistent with the properties of the main star-forming regions (A, B and C) in Sextans A \citep{vDPW98, Dohm02} . We find most young stars concentrated in regions B and C, coincident with main galactic overdensities of neutral hydrogen, similarly to our findings in IC1613 \citep{GH10}.

\begin{acknowledgements}

Funded by Spanish MINECO under grants AYA-2012-39364-C02-01, AYA-2010-21697-C05-01, ESP2013-47809-C3-1-R, FIS2012-39162-C06-01 and SEV-2011-0187, and by Gobierno de Canarias under grant PID2010119. This research has made use of Aladin \citep{aladin}, NASA'a Astrophysics Data System and the IAC Supercomputing facility HTCondor (http://research.cs.wisc.edu/htcondor/).

\end{acknowledgements}


\Online 
 \begin{appendix}

\section{Notes on individual targets}         
\label{SS:NiT}

\textbf{\textit{OB326 (O7.5 III ((f)))}:} The spectral SNR is poor as expected considering the V$\sim$ 20.7 magnitude of the target. Yet, the detection of He\,{\sc II} 4541, He\,{\sc II} 4686 and He\,{\sc I} 4471 is clear. The nebular extraction at the Balmer lines is incomplete, but it may have been oversubtracted in other lines (i.e. He\,{\sc II} 4200 and He\,{\sc I} 4471). 

The He\,{\sc II} 4541/He\,{\sc I} 4471 ratio suggests O7-O8 type, but if  He\,{\sc I} 4471 had been indeed oversubtracted, the spectral type could be earlier. The unclear detection of He\,{\sc I} 4387 and He\,{\sc I} 4144 also supports an earlier type. On the other hand, the detection of Si\,{\sc IV} 4089  suggests a spectral type later than O7.  We adopt O7.5 as a compromise. The width of the Balmer lines suggest luminosity class \,{\sc III}. Since there may be emission of N\,{\sc III} 4634-40-42, our final classification is O7.5\,{\sc III} ((f)).

\textbf{\textit{OB623 (O8 Ib)}:} The SNR is poor as expected for a star with V-magnitude $\sim$ 20.6 in a dense nebular region. The pre\-sence of  He\,{\sc II} 4541 in the spectra indicates an O-type star, and the absence of Si\,{\sc III} lines points to a spectral type earlier than O9. Assuming that the identification of the He\,{\sc II} 4541, He\,{\sc I} 4388 and He\,{\sc I} 4471 lines are correct, their ratios suggest an O8 type. 
There is no emission or absorption of He\,{\sc II} 4686, so the final classification is O8 Ib.

\textbf{\textit{OB521 (O9.5 III-V)}:} The He\,{\sc II} lines are still visible in this spectrum together with He\,{\sc I} lines, which indicates a late O-type.
All the Sota et al.' spectral diagnostic criteria suggest an O9.5 star. The He\,{\sc II} 4686/He\,{\sc I} 4713 and Si\,{\sc IV} 4089/He\,{\sc I(+II)} 4026 ratios disagree, although we favor the first one that does not depend on the Si abundance, and we adopt  luminosity class \,{\sc III-V}.  
A nearby star, with V-magnitude $\sim$ 19.3 and redder color (according to Massey's photometry), could be contaminating the spectrum of this star.

\textbf{\textit{OB523 (O9.7 I((f)))}:} The He\,{\sc II} 4541/He\,{\sc I} 4471 ratio suggests spectral type O8, but the preferred Sota et al. He\,{\sc II} 4541/He\,{\sc I} 4387 criterion suggests spectral type O9.7, which we adopt. 
There is no clear detection of the additional diagnostic lines He\,{\sc II} 4200 or He\,{\sc I} 4144.  The Si\,{\sc III} 4552 line is weaker than expected for the O9.7 types, but this could be due to the low metal content of the galaxy.
As for Sota et al.'s luminosity diagnostic lines, the He\,{\sc II} 4686/He\,{\sc I} 4713 relative strength suggests a supergiant class Ib.  There is a small emission of N\,{\sc III} 4634-40-42. The final classification is O9.7 Ib((f)).

This star is contaminated by a nearby star which has a similar V-filter magnitude and redder color, according to Massey's photometry. The latter is weaker in the B-filter by approximately two magnitudes and slightly displaced from the center of the slit, which implies only a minor contamination to the blue spectrum of OB523. This star also presents an anomalously high radial velocity. We do not detect any nearby bow-shock structure in the optical images consistent with a recent ejection.

\textbf{\textit{OB321 (O9.7 I((f)))}:} The He\,{\sc II} lines in the spectrum of OB321 are clear, but weak compared to the He\,{\sc I} lines. Their relative strength indicate subtype O9.7. 
Sota et al.'s diagnostic He\,{\sc II} 4686/He\,{\sc I} 4713, and the narrow observed Balmer lines,  point to a luminosity class \,{\sc I}. 
N\,{\sc III} 4634 exhibits a small emission.

\textbf{\textit{OB622 (B0 I)}:} No He\,{\sc II} lines are detected, but the He\,{\sc I} lines are strong and clear indicating a B-type. Mg\,{\sc II} 4481 is weak, suggesting an early B-type, possibly B0. He\,{\sc I} 4713 is visible while He\,{\sc II} 4686 is absent, and we assign luminosity class \,{\sc I}.

\textbf{\textit{OB524 (B0 III)}:}  The SNR is poor, as expected considering the star's V-magnitude. OB524 is located in an intense nebula, with nearby stars that could also contaminate the observations. The spectrum displays He\,{\sc I} absorptions, but the lines of Si\,{\sc III} and Mg\,{\sc II} are not clearly detected, indicating an early B-type. 
The observed strong emission of He\,{\sc II} 4541 and the P~Cygni-like He\,{\sc II} 4686 are both artifacts introduced during the background subtraction. 

The Balmer lines are broad, with overlapped signatures of sky oversubtraction. We classify the star as B0\,{\sc III}.

\textbf{\textit{OB222 and OB421 (B1 I)}:} These stars show no He\,{\sc II} lines in their spectrum (except for a trace of He\,{\sc II} 4686) but the He\,{\sc I} lines are clear. Mg\,{\sc II} 4481 is absent, indicating B0-B2 types.
We note that OB222 shows a strong absorption close to Mg\,{\sc II} 4481 but its center does not match the central wavelengths of the Mg\,{\sc II} transition. Since the Si\,{\sc III} lines (4552, 4567 and 4574 \AA) are strong, we assign spectral type B1.

The supergiant classification is inferred from Balmer lines and the Si\,{\sc IV} 4089/He\,{\sc I} 4121 ratio.
The sky has been oversubtracted in the OB421 spectrum.

Both stars present anomalous radial velocity, lower than the galaxy's curve for OB222 and higher for OB421. The optical images do not provide additional clues on their possible runaway nature.

\textbf{\textit{OB422 (B1 I)}:} The He\,{\sc I} lines are strong. Mg\,{\sc II} 4481 is weak but clear, however Si\,{\sc III} 4552 is not clearly detected. The strong Si\,{\sc IV} 4089 line indicates  luminosity class \,{\sc I}. 

OB422 is located in a region with nebulosity and this reflects on a poor background subtraction.

\textbf{\textit{OB323 (B1 III)}:} This star exhibits strong lines of He\,{\sc I} and Si\,{\sc III}. Because Mg\,{\sc II} 4481 is weak we classify the star as B1.

We adopt luminosity class \,{\sc III} because of the width of $H_{\beta}$.

\textbf{\textit{OB221 (B2.5 I)}:} Its spectral features suggest contradicting classifications. While the weak Mg\,{\sc II} 4481 indicates an early B-type, the silicon features (absence of Si\,{\sc IV} 4089 and Si\,{\sc III} 4552 and detection of Si\,{\sc II} 4128) point to a later type. We adopt B2.5 I as a compromise, the luminosity class assigned based on the width of the Balmer lines.

\textbf{\textit{OB324 and OB121 (B8 II)}:} The strong Mg\,{\sc II} 4481 line, as deep as He\,{\sc I} 4471, indicates  B8 type. The presence of He\,{\sc II} 4686 emission in the OB324 spectrum is possibly related to a cosmic ray. The width of the Balmer lines su\-ggests  luminosity class \,{\sc II}.
The presence of [O{\sc III}] absorption lines in OB121 indicates an oversubtraction of the sky.

OB121 exhibits lower SNR than expected from its V-magnitude and exposure time, as a consequence of a number of observing issues. OB121 was included in slit-1 and slit-5. Because it included brighter targets, the exposure time allocated to slit-1 is shorter than for other configurations (see Table \ref{T:log}). On the other hand, OB121 is off-slit in slit-5. Finally, both slit-1 and -5 were observed at larger airmass than the rest of the observations.

\textbf{\textit{OB122 (B9 I)}:}  This star still displays He\,{\sc I} lines. The intense Si\,{\sc II} 4128 line and the absence of higher ionization stages of Si, indicates a late-B type. Since Mg\,{\sc II} 4481 is deeper than He\,{\sc I} 4471 we classify the star as a B9 supergiant, the luminosity class determined by the width of the Balmer lines.

\textbf{\textit{OB525 (A0 II)}:} The star lacks spectral lines of He\,{\sc II} or He\,{\sc I} but the Balmer lines are strong, indicating A-type or later. The spectrum does not show the G-band at 4300 \AA~ but it displays Fe\,{\sc II} and Ca\,{\sc I} lines, which leads to an A0 type. 
We assign luminosity class \,{\sc II} based on the equivalent width of the Balmer lines.

The [O{\sc III}] lines in absorption and the core of the Balmer lines indicate oversubtraction of nebular lines. OB525 is not hot enough to ionize gas, but the star is located close to the H\,{\sc II} bubbles of region-B.

\textbf{\textit{OB625 (A5 II)}:} The poor SNR of its spectrum hinders its spectral classification. Because the Balmer lines are strong, the star is earlier than F5. The spectral features in the 4000-5000~\AA~ range resemble an A0-type star, but the 5000-5500~\AA~ range (not shown) rather suggest A5-F5 types. We adopt A5 as a compromise.

The luminosity class {\sc II} was assigned based on the Balmer series width.

\textbf{\textit{OB621 (F5 I)}:} The strength of the G-band relative to the Fe\,{\sc I} + Fe\,{\sc II} 4383-85 and Ca\,{\sc I} 4226 lines yields spectral type F5. Since the Fe\,{\sc II} 4173-78 lines are absent, we assign luminosity class \,{\sc I}.

OB621 is in region-B, surrounded by nebulosity; in fact, the emission of the [O{\sc III}] lines indicate an insufficient background subtraction.

\citet{KVT04} classified OB621 as an F-hypergiant. One of the goals of this paper was to check for spectral type variations that would make the star an LBV-candidate. Our F5 type concurs with the F classification, however, we have found luminosity class {\sc I}. We have compared the spectral morphology of OB621 with the F-hypergiant B324 in M33 \citep{Monteverde96}. While OB621 exhibits absorption hydrogen lines, the Balmer series is in strong emission in B324 supporting our classification of OB621 as a supergiant. Since \citet{KVT04} 's spectrum is not available, we cannot assess the extent and implications of the change in luminosity class.

 \end{appendix}

\end{document}